# Quantitatively Analyzing Phonon Spectral Contribution of Thermal Conductivity Based on Non-Equilibrium Molecular Dynamics Simulation II: From Time Fourier Transform

Yanguang Zhou[1] and Ming Hu[1, 2, *]

[1]*Aachen Institute for Advanced Study in Computational Engineering Science (AICES), RWTH Aachen University, 52062 Aachen, Germany*

[2]*Institute of Mineral Engineering, Division of Materials Science and Engineering, Faculty of Georesources and Materials Engineering, RWTH Aachen University, 52064 Aachen, Germany*

## Abstract

From nano-scale heat transfer point of view, currently one of the most interesting and challenging tasks is to quantitatively analyzing phonon mode specific transport properties in solid materials, which plays vital role in many emerging and diverse technological applications. It has not been so long since such information can be provided by the phonon spectral energy density (SED) or equivalently time domain normal mode analysis (TDNMA) methods in the framework of equilibrium molecular dynamics simulation (EMD). However, until now it has not been realized in non-equilibrium molecular dynamics simulations (NEMD), the other widely used computational method for calculating thermal transport of

---

materials in addition to EMD. In this work, a computational scheme based on time Fourier transform of atomistic heat current, called frequency domain direct decomposed method (FDDDM), is proposed to analyze the contributions of frequency dependent thermal conductivity in NEMD simulations. The FDDDM results of Lennard-Jones (LJ) Argon and Stillinger-Weber (SW) Si are compared with TDNMA method from EMD simulation. Similar trends are found for both cases, which confirm the validity of our FDDDM approach. Benefiting from the inherent nature of NEMD and the theoretical formula that does not require any temperature assumption, the FDDDM can be directly used to investigate the size and temperature effect. Moreover, the unique advantage of FDDDM prior to previous methods (such as SED and TDNMA) is that it can be straightforwardly used to characterize the phonon frequency dependent thermal conductivity of disordered systems, such as amorphous materials. The FDDDM approach can also be a good candidate to be used to understand the phonon behaviors and thus provides useful guidance for designing efficient structures for advanced thermal management.

# I. INTRODUCTION

The conductive heat transfer in a solid or a quiescent fluid can be described by Fourier's law [1]. Essentially, the thermal conductivity is related to the properties of the sub-continuum energy carrier. Generally speaking, in solids the energy carriers are electrons and phonons, i.e., lattice vibrations. The electrons are responsible for the majority of heat conduction in metal, while the phonons are the dominant carriers in the semiconductors and insulators [2], e.g. solid argon, silicon, germanium and silica-based materials. Here, we restrict our focus on the semiconductors and insulators, in which the thermal conductivity is mainly dominated by the phonons. Understanding the modal contribution to thermal conductivity can be quite important for studying thermal transport properties of many nanotechnology devices [3]. Many powerful methods [4-8] were generated to obtain the modal contribution of phonons. All these methods can deal with relatively simple systems very well such as perfect bulk crystalline materials, while for some complicated systems such as systems with defects and interfaces, these methods can only give the approximate solution with some assumptions.

From physics point of view, the thermal transport at small scales is quite different from that at macro-scales [1, 9]. Some formidable approaches [10-13] have been proposed from the phonon level or atomistic level. Recently, Broido *et a.l* [14], Li *et al.* [15] and Chernatynskiy *et al.* [16] use the force constants calculated by first principles to generate the input parameters needed for Boltzmann transport equation (BTE) via lattice dynamics, and give more accurate thermal conductivity of materials. However, such method only considers the three-phonon scattering process so far. The classical molecular dynamics (MD) method, on the other hand, can describe the phonon-phonon interactions at all levels exactly. In the framework of MD, two popular ways, namely Green-Kubo (GK-MD) method [17, 18] and the non-equilibrium molecular dynamics (NEMD) [19-21], are widely adopted to calculate the thermal conductivity. However, for a rather long time both methods can only predict the overall thermal conductivity rather than provide more information of phonon properties, such as the mode specific thermal conductivity. Recently, the spectral energy density (SED) [22-27] or equivalently the time domain normal mode analysis (TDNMA) [28-33] methods based on EMD simulations and the time domain direct decomposition method (TDDDM) [34, 35] in the framework of NEMD are proposed to calculate the mode specific thermal conductivity by directly evaluating the phonon lifetime for each single mode. However, for the methods based on EMD, they can tackle the relatively simple systems (such as bulk crystalline) successfully, but for complicated systems, e.g. disordered materials, some assumption in calculating phonon group velocity has to be made when SED or TDNMA is used [32, 36]. For TDDDM based on NEMD [34], since we should guarantee the heat current of all phonon modes to reach steady state, the computational cost of TDDDM is quite high for large system.

In this paper, a new method named frequency domain direct decomposition method (FDDDM), which is still based on NEMD simulations, is proposed from the atomistic heat current formula to calculate the spectrum contribution to thermal conductivity. The effects of all level phonon scattering processes are included in our method, since we do not neglect the higher-order phonon scattering in simulations. Using our approach, it is straightforward to find that the number of phonons in the low frequency range, or equivalently with long mean free path (MFP), increases with system size increasing. It is also straightforward to identify that the high temperature enhances the anharmonic energy transfer in the system, which is well known and proves that our FDDDM reveals the correct physical picture of phonon-phonon scattering. Furthermore, the most striking feature of this new method is to characterize the phonon properties of complicated systems, such as amorphous or alloyed materials. Unlike the method in Ref. [37] which only considers the first order anharmonic interactions between two atoms, our new FDDDM approach includes all orders of anharmonicity in the interatomic interaction.

The remainder of this paper is organized as follows. In Sec. II, the theoretical framework of our FDDDM method is presented to characterize the phonon mode contribution to the overall thermal conductivity. Then the implementation details of the FDDDM scheme are given in Sec. III and the validation of the frequency dependent thermal conductivity is given in Sec. IV, followed by the discussions on some important computational parameters in the FDDDM framework in Sec. V. In Sec. VI, we apply FDDDM to study the size and temperature effect on the frequency dependent phonon mode contribution in perfect crystalline systems (LJ Argon and SW Si). In addition, phonon transport in amorphous Si as a representative of disordered systems is studied by FDDDM. In Sec. VII, we present a summary and conclusion.

## II. THEORY OF CHARACTERIZING SPECTRAL CONTRIBUTION TO THERMAL CONDUCTIVITY

It is well known that, NEMD simulations explicitly rely on launching a steady heat flux flowing through the system. The atomistic heat current $\mathbf{q}_{ij}$ between two atoms $i$ and $j$ can be written as [37-39]

$$\mathbf{q}_{ij} = \frac{1}{2}\langle \mathbf{F}_{ij}(\mathbf{v}_i + \mathbf{v}_j)\rangle, \qquad (1)$$

where $\mathbf{F}_{ij}$ is the force between two atoms and $\mathbf{v}$ is the velocity of atoms, $\langle\ \rangle$ denotes the time average. Meanwhile, it has been proved that the spectrum decomposition of heat current is related to the correlation time between force and velocity [37, 38]. Therefore, we can write the auxiliary correlation term $\mathbf{C}_{ij}$ as

$$\mathbf{C}_{ij}(t_2 - t_1) = \langle \mathbf{F}_{ij}(t_1) \cdot (\mathbf{v}_i(t_2) + \mathbf{v}_j(t_2)) \rangle, \qquad (2)$$

where $\mathbf{C}_{ij}$ only depends on the time difference due to the stable state and ensemble average of the system.

By defining characteristic time $\tau$ as

$$\tau = t_1 - t_2, \qquad (3)$$

we can re-write the atomistic heat current in the form of

$$\begin{aligned}
\mathbf{q}_{ij} &= \frac{1}{2}\lim_{t_2 \to t_1}\left\langle \mathbf{F}_{ij}(t_2)\cdot(\mathbf{v}_i(t_1)+\mathbf{v}_j(t_1))\right\rangle \\
&= \frac{1}{2}\lim_{\tau \to 0}\left\langle \mathbf{F}_{ij}(\tau)\cdot(\mathbf{v}_i(0)+\mathbf{v}_j(0))\right\rangle \\
&= \frac{1}{2}\lim_{\tau \to 0}\mathbf{C}_{ij}(\tau)
\end{aligned} \qquad (4)$$

As usual, we can define the time Fourier transform $\widetilde{C}_{ij}(\omega) = \int_{-\infty}^{+\infty} C_{ij}(\tau)e^{i\omega\tau}d\tau$. Therefore, $C_{ij}(\tau)$ can be obtained by performing the inverse transform of $\widetilde{C}_{ij}(\omega)$, $C_{ij}(\tau) = \int_{-\infty}^{+\infty}\widetilde{C}_{ij}(\omega)e^{-i\omega\tau}\frac{d\omega}{2\pi}$. Here, the $\omega$ represents the angular frequency. Thus, $q_{ij}$ can be further given by

$$q_{ij} = \frac{1}{2}\lim_{\tau \to 0}\int_{-\infty}^{+\infty}\widetilde{C}_{ij}(\omega)e^{-i\omega\tau}\frac{d\omega}{2\pi} = \frac{1}{2}\int_{-\infty}^{+\infty}\widetilde{C}_{ij}(\omega)\frac{d\omega}{2\pi} = \frac{1}{2}\int_{-\infty}^{+\infty}\int_{-\infty}^{+\infty}C_{ij}(\tau)e^{i\omega\tau}d\tau\frac{d\omega}{2\pi} \ . \qquad (5)$$

Since $C_{ij}(\tau)$ is a real number, the real and imaginary part of $\widetilde{C}_{ij}(\omega)$ should be an even and odd function. Combining with Eq. (4), the spectrum of atomistic heat current can be obtained by

$$q_{ij}(\omega) = \frac{1}{2}\int_{-\infty}^{+\infty} C_{ij}(\tau)e^{i\omega\tau}d\tau = \text{Re}\left[\int_0^{+\infty} C_{ij}(\tau)e^{i\omega\tau}d\tau\right] \ . \qquad (6)$$

On the other hand, the molecular dynamics expression of the heat current, which was first derived by Irving and Kirkwood[40], and later on extended by Daichi et al.[41, 42], can be written as[43]

$$Q = \sum_i \langle E_i v_i \rangle + \sum_n \sum_{S_1}\sum_{S_1<S_2}\cdots\sum_{S_{n-1}<S_n}\left[\sum_{\alpha=1}^{n-1}\sum_{\beta=\alpha+1}^{n}\langle F_{S_\alpha S_\beta}\cdot\left(P_{S_\alpha}^n v_{S_\alpha} + P_{S_\beta}^n v_{S_\beta}\right)\cdot\left(r_{S_\alpha} - r_{S_\beta}\right)\rangle\right] ,$$
$$(7)$$

where $\mathbf{Q}$ is the heat current, the sum is taken over all atoms which are described by $n$-body potential in the control volume, $s_i$ denotes the atoms, $E_i$ and $\mathbf{r}_i$ are the energy and position of atoms, respectively. $P$ is the distribution of the interaction potential and satisfies $\sum_n P_i = 1$ [41]. In the classical systems, equal partition of potential energy between atoms $P_i^n = 1/n$ is usually assumed [41].

Since the small fluctuations in the atomic coordinates cannot cause energy transfer over macroscopic distance and time, there is negligible macroscopic atom diffusion in a solid material [28, 44]. Then, for solid materials the Eq. (7) can be simplified as

$$\mathbf{Q} = \sum_n \sum_{S_1} \sum_{S_1 < S_2} \cdots \sum_{S_{n-1} < S_n} \left[ \sum_{\alpha=1}^{n-1} \sum_{\beta=\alpha+1}^{n} \langle \mathbf{F}_{S_\alpha S_\beta} \cdot \left( P_{S_\alpha}^n \mathbf{v}_{S_\alpha} + P_{S_\beta}^n \mathbf{v}_{S_\beta} \right) \cdot \left( \mathbf{r}_{S_\alpha} - \mathbf{r}_{S_\beta} \right) \rangle \right].$$

(8)

Furthermore, since for a stable system $\left( r_{S_\alpha}^0 - r_{S_\beta}^0 \right) \gg \left( u_{S_\alpha} - u_{S_\beta} \right)$, which has already been proved by our MD simulations (see Appendix A for details), the position difference of the two atoms $(\mathbf{r}_{S_\alpha} - \mathbf{r}_{S_\beta})$ can be assumed as

$$\begin{aligned} (\mathbf{r}_{S_\alpha} - \mathbf{r}_{S_\beta}) &= (\mathbf{r}_{S_\alpha}^0 - \mathbf{r}_{S_\beta}^0) + (\mathbf{u}_{S_\alpha} - \mathbf{u}_{S_\beta}) \\ &\approx (\mathbf{r}_{S_\alpha}^0 - \mathbf{r}_{S_\beta}^0) \end{aligned}, \quad (9)$$

where $\mathbf{r}_i^0$ and $\mathbf{u}_i$ represents the equilibrium position and displacement of atom $i$.

Using Eqs. (4), (5), (7), (8) and (9), the heat current can be written as

$$\mathbf{Q} = \sum_n \sum_{S_1} \sum_{S_1 < S_2} \cdots \sum_{S_{n-1} < S_n} \left[ \sum_{\alpha=1}^{n-1} \sum_{\beta=\alpha+1}^{n} \frac{1}{n} \mathbf{q}_{S_\alpha S_\beta} \cdot \left( r_{S_\alpha}^0 - r_{S_\beta}^0 \right) \right]$$

$$= \sum_n \sum_{S_1} \sum_{S_1<S_2} \cdots \sum_{S_{n-1}<S_n} \left[ \sum_{\alpha=1}^{n-1} \sum_{\beta=\alpha+1}^{n} \frac{1}{2n} \int_{-\infty}^{+\infty} \int_{-\infty}^{+\infty} C_{ij}(\tau) e^{i\omega\tau} d\tau \frac{d\omega}{2\pi} \cdot \left( r_{S_\alpha}^0 - r_{S_\beta}^0 \right) \right].$$
(10)

And the spectrum of heat current in a control volume can be obtained as

$$\boldsymbol{Q}(\omega) = \sum_n \sum_{S_1} \sum_{S_1<S_2} \cdots \sum_{S_{n-1}<S_n} \left[ \sum_{\alpha=1}^{n-1} \sum_{\beta=\alpha+1}^{n} \frac{1}{n} \text{Re} \left[ \int_0^{+\infty} C_{ij}(\tau) e^{i\omega\tau} d\tau \right] \cdot \left( r_{S_\alpha}^0 - r_{S_\beta}^0 \right) \right].$$
(11)

Finally, by assuming the same temperature gradient for all phonon modes, the contribution of phonon mode with a specific frequency to the overall thermal conductivity can be calculated by Fourier's law

$$\mathbf{K}(\omega) = -\frac{1}{V_c} \frac{\boldsymbol{Q}(\omega)}{\nabla T},$$
(12)

where $V_c$ is the volume of control box in a NEMD simulation (see Fig. 1 in Sec. III for details). Note that $\mathbf{K}(\omega)$ is a 3 × 3 tensor, since $\nabla T$ be along either the same or different direction as the heat flux $\boldsymbol{Q}(\omega)$.

Using the relaxation time approximation[45] to solve Boltzmann transport equation yields an expression of modal thermal conductivity $\mathbf{K}_{\text{BTE}}(\mathbf{k}, \nu)$ [46]

$$\mathbf{K}_{\text{BTE}}(\mathbf{k}, \nu) = c_{ph}(\mathbf{k}, \nu) \mathbf{v}_g^2(\mathbf{k}, \nu) \tau(\mathbf{k}, \nu).$$
(13)

The modal relaxation time $\tau(\mathbf{k}, \nu)$ can be calculated using TDNMA (details can be found in our previous work [34]). The group velocity $\mathbf{v}_g(\mathbf{k}, \nu)$ can be obtained from lattice dynamics and the volume specific heat $c_{ph}(\mathbf{k}, \nu)$ can be calculated by $c_{ph}(\mathbf{k}, \nu) = k_b / V$, in which $k_b$ and V is the Boltzmann constant and system volume, respectively.

It is well known that for each frequency $\omega$, there are typically multiple phonon modes with different polarizations and vectors (**k**, ν). In order to compare the results between FDDDM and TDNMA directly, we compute thermal conductivity contributed by each frequency in TDNMA using

$$\mathbf{K}_{\mathrm{BTE}}(\omega) = \sum_{k,v} \mathbf{K}_{\mathrm{BTE}}(\omega; k, v) , \qquad (14)$$

where the thermal conductivity of a specific frequency is taken over different branches (e.g. LA, TA, TO and LO) for the same frequency without using broadening functions.

Meanwhile, we also generated the modal thermal conductivity from the space Fourier transform which is called TDDDM in our first paper in the series. Here we give some comparisons of these two methods from the theory aspect. For FDDDM we start from the auxiliary correlation between atomic force (***F***) and velocity (***v***) and then obtain the heat current spectrum. For TDDDM we directly get the heat current of each mode from ***F*** and the decomposed ***v***. Although these two approaches are fundamentally different (the FDDDM is based on the Fourier transform of time and the TDDDM is from the Fourier transform of space), we speculate that there might be some relations between the time and space Fourier transform. However, it is really beyond the authors' knowledge to prove this theoretically. Although it is natural to obtain similar results using these two methods, so far no one can prove these formulas are equivalent. It is well known that the thermal conductivity in semiconductor is mainly from lattice vibrations. Here, the result of our FDDDM is generated from the atomic heat current which only considers the phonons' contribution (no effects of electric and radiation). Therefore, our results do not include an arbitrary frequency which is corresponding to electrons or others.

### III. COMPUTATIONAL IMPLEMENTATION OF FDDDM

The FDDDM proposed above is based on the decomposition of heat current in non-equilibrium molecular dynamics simulations. The NEMD setup is depicted schematically in Fig. 1. The approach to generate heat current, the boundary conditions and interaction potential are the same as that in our previous paper [34]. Simulations with different temperatures (10 – 40 K for LJ Argon and 300 – 600 K for SW Si) are implemented, and therefore are used to investigate the temperature effect on the spectral contribution to thermal conductivity. On the other hand, as we discussed in our first paper [34], there always exist size effects for thermal conductivity in NEMD simulations [19, 47]. Here, we choose SW Si as an example to analyze the size effects in NEMD simulations via FDDDM. To this end, the simulation systems are chosen as $8a_{Ar} \times 8a_{Ar} \times 200a_{Ar}$ for LJ Argon and $8a_{Si} \times 8a_{Si} \times n \cdot a_{Si}$ for SW Si (*a* is the lattice parameters, i.e. 0.529 nm and 0.544 nm for LJ Argon and SW Si, respectively), where *n* ranges from 50 to 1600. The ensemble used for our system and the way to obtain the steady heat current are similar to that in our first paper [34]. The temperature gradient $\nabla T$ is guaranteed to be not too large (maximum 0.025 K/$a_{Ar}$ and 0.5 K/$a_{Si}$ for LJ Argon and SW Si, respectively), so that the response of the system to the heat flux can be kept in the linear regime. Finally, the frequency dependent heat current is extracted from the additional 5 ns simulations. The time steps in all our simulations are 1 fs.

Another important thing in FDDDM is the selection of implemented volume which is called "control volume" in our first paper [34] in our simulations. Apart from the distance between the control volume and the thermostat reservoir and the computational cost as we discussed before, we also need to consider the size of control volume since large number of atoms can ensure the inherent temperature fluctuation can be ignored in our simulations. Therefore, the size of the control volume is usually the compromise between enough large number of atoms contained in the control volume and fast computation. In our simulations, the size of the control volume of LJ Argon and SW Si is $8a_{Ar} \times 8a_{Ar} \times 4a_{Ar}$ and $8a_{Si} \times 8a_{Si} \times (3-8)a_{Si}$, respectively. The corresponding total number of atoms in the control volume are 1024 and 1536 – 4096 for LJ Argon and SW Si, respectively. We have checked that all the control volumes in our simulations are located in the linear region of temperature gradient and the number of atoms in control volume is large enough to make sure the intrinsic temperature fluctuation can be ignored.

In order to validate the results of our FDDDM approach, the TDNMA technique based on the EMD simulations is used to obtain the phonon behaviors. All details about TDNMA simulation can be found in our previous work [34]. As we discussed before, the size of FDDDM models in the concerned direction should be larger than 65 nm for LJ Argon and 2 μm for SW

Si, if all the phonon modes appearing in our TDNMA model can be included in FDDDM.

## IV. VALIDATION OF FREQUENCY DOMAIN DIRECT DECOMPOSITION METHOD

In this section, the frequency dependent thermal conductivity contribution is predicted by FDDDM and the result is compared with that obtained by TDNMA. Meanwhile, based on Sec. II and our previous work [34], the size of the FDDDM models mentioned here should be at least $8a_{Ar} \times 8a_{Ar} \times (> 123a_{Ar})$ for LJ Argon and $8a_{Si} \times 8a_{Si} \times (> 3676a_{Si})$ for SW Si. Here, due to the computational efficiency, we choose the size of FDDDM models as $8a_{Ar} \times 8a_{Ar} \times 200a_{Ar}$ for LJ Argon and $8a_{Si} \times 8a_{Si} \times 2000a_{Si}$ for SW Si. Thus, for LJ Argon all extractable phonon modes in TDNMA can be included in FDDDM simulations, while for SW Si about 10% of phonon modes that can be extracted in TDNMA are truncated (ignored) in the FDDDM simulations. Furthermore, the full Brillouin zone, which has been proved to be more appropriate than the isotropic approximation, is used to predict the thermal conductivity in TDNMA [26, 30, 46, 33]. In this paper, in order to improve the computational efficiency, the crystal lattice's irreducible Brillouin zone is used to compute the contribution of each phonon mode, which has been shown to predict the same thermal conductivity of a material with respect to use the full Brillouin zone [33]. Figure 2 shows the results predicted by FDDDM and TDNMA. For both LJ Argon and SW Si, the results obtained by FDDDM show similar trends with that calculated by using TDNMA.

For LJ Argon at 10K (left panel of Fig. 2), the results show that our FDDDM approach yields the similar results with TDNMA. It is not surprising to see this phenomenon, since the largest MFP appearing in our TDNMA simulations is smaller than the size of the system in FDDDM. Therefore, all the extractable phonon modes in TDNMA are present and extracted in our FDDDM simulation. Meanwhile, the phonon modes in the medium frequency region (0.7 – 1.6 THz) are found to dominate the phonon transport (contribute more than 90% of the total thermal conductivity). The phonon modes in the low frequency region (< 0.7 THz) and high frequency region (> 1.6 THz) contribute very little to the total thermal conductivity. All our findings are consistent with that observed in Ref. [ 48].

For SW Si at 300 K (right panel of Fig. 2), the predictions of our FDDDM method is slightly different from that of TDNMA. We attribute the difference to the size effects in FDDDM, since it is based on NEMD. To further prove this, the size along $z$ (heat flux) direction in the system is reduced to be 425 nm. It is clearly seen that the results of FDDDM for the longer system is much closer to the predictions of TDNMA, in particular the accumulative thermal conductivity in the low frequency range has a slight augment. Therefore, we anticipate that the results calculated by FDDDM will be in accordance with the prediction of TDNMA when the largest extractable MFP in TDNMA can be included in the system of FDDDM simulations. Furthermore, from Fig. 2 we can see that the acoustic phonon modes ($\omega < 12$ THz for SW Si) contribute about 90% of the predicted thermal conductivity. Thus, it is reasonable to conclude that the optical phonon modes ($\omega > 12$ THz for SW Si) do not contribute much to the overall thermal conductivity [48, 49]. In addition, phonon modes in low frequency region (< 8 THz) are found to dominate the thermal conductivity (the accumulative thermal conductivity reaches about 80% for frequency up to 8 THz). Both previous work [34, 49] and this work have shown that, the common assumption that low frequency phonon modes dominate the thermal transport [1, 9] should be considered cautiously (invalid for LJ Argon, while valid for SW Si).

Although the relative contribution of each phonon mode should be the same when there is no size effect in the system, since the same phonon should have the same effect in the same system, the actual thermal conductivities calculated using TDNMA/SED (EMD) and FDDDM (NEMD) are different (insets in Fig. 2). The discrepancy may come from the size effect (such as SW Si in Fig. 2) and more importantly the inherent difference between EMD and NEMD. Thus, it is difficult to directly compare the actual thermal conductivity between FDDDM and SED/TDNMA.

## V. IMPORTANT PARAMETERS IN FDDDM SIMULATIONS

### A. Run time and computational cost of FDDDM

As mentioned above, the FDDDM is based on NEMD simulation. Before the FDDDM calculation is switched on, one should run enough long time to build the stable temperature gradient in the system. In our simulations, 3 – 5 ns (LJ Argon) and 5 – 10 ns (SW Si) are used to build temperature gradient in the concerned direction. Meanwhile, for FDDDM the heat current spectrum is related to the autocorrelation between the force and velocity of atoms (Eq. [6]). Therefore, the run time of FDDDM simulations should be also long enough such that all the autocorrelations (Eq. [2]) can converge. It is also worth noticing that the autocorrelation in Eq. (2) is very sensitive to the oscillation of heat current. Therefore, the last 2 ns data in the FDDDM simulations are used to obtain the final heat current spectrum by using the Bartlett's method[50]. As discussed in our previous work[34], in TDDDM it takes a quite long time to let all the phonon modes reach steady state (about 6 ns for LJ Argon and > 30 ns for SW Si), while for FDDDM the run time which is needed to obtain the stable heat current spectrum is much shorter (4 ns for LJ Argon and SW Si). The reason is that, it is easier to catch the phonon interactions (autocorrelation in Eq. [2] in this paper) than to depict the variation of individual phonon mode (Eq. [9] in Ref. [34]). Phonons with large relaxation time, which reaches about 100 ps for LJ Argon and over 600 ps for SW Si, are very difficult to be caught in TDDDM, since these rare scattering events need very long time to occur.

The computational cost of our FDDDM method is also a major concern. The TDNMA method scales as $[(Nn)^2]$, where $N$ and $n$ represents the number of unit cell and basis atoms in the unit cell, respectively. In contrast, the computational cost of FDDDM method is proportional to $[(Nn)]$ since $\boldsymbol{C}(\tau)$ is computed by $Nn$ times, in comparison to that of TDDDM which scales as $2[(Nn)^2]$. Obviously the cost of pure FDDDM approach is much smaller. This enables us to use the FDDDM method to tackle relatively large systems.

However, since both FDDDM and TDDDM are based on NEMD, the system should run a long time to make the heat current of the system reach the steady state. Therefore, the total running time for both FDDDM and TDDDM are much larger than TDNMA/SED. Another important issue is the storage space. Since both methods mentioned above rely on post-processing, a quite large space (over 10 GB) is required to save the data throughout the production run [23]. Here, we embedded the TDNMA and FDDDM methods into the open source package LAMMPS [51], and therefore the required space to save data, without storing the intermediate force and velocity information, is reduced to hundreds of MBs.

### B. Effect of the size of the system and control volume

The size effect cannot be avoided in FDDDM, since it is directly based on the NEMD simulations. As discussed in our previous work [23], the phonon modes with MFP larger than system size will be truncated. For LJ Argon, the largest MFP of phonons is only several hundred of nanometers [48]. Thus, the size effects of LJ Argon can be ignored, since it is easy to include almost all the phonon in our FDDDM simulations, while for SW Si the largest MFP can be as large as several micrometers [49], which is unreachable for NEMD simulations. Strong size effects can be found when our FDDDM approach is implemented into the SW Si systems (Fig. 3). The details of the size effect in FDDDM are given in the next section.

From Sec. III we know that the size of the control volume may also affect the results of FDDDM simulations. Other than TDNMA and TDDDM in our previous work [34], where the results of which are related to the system size (control volume in TDDDM), FDDDM is essentially independent of the size of the control volume, since it is derived from Fourier transform of time rather than space (Eq. [6]). However, as we mentioned in Sec. II, more atoms in the control volume will decrease the fluctuation of heat current. Here, the size of the control volume is chosen to be $4.2 \times 4.2 \times 4.2$ nm$^3$ for LJ Argon and $1.6 \times 4.4 \times 4.4$ nm$^3$ for SW Si. We also enlarge the size of the control volume by 2 times larger and find that the results of FDDDM simulations are almost the same (difference less than 0.3%). It is also worth pointing out that, there is a problem of finite phonons can be extracted in the control volume when using TDDDM and we can avoid such problem by using FDDDM. Unlike TDDDM, all the phonons can be captured by FDDDM. For TDDDM which is based on lattice dynamics, we do the space Fourier transform and then the number of extractable phonons is limited, since it is determined by the number of atoms in lattice dynamics. In contrast, for FDDDM we do the time Fourier transform and then the number of extractable phonons is infinite. Thus, the control volume should be large enough in order to obtain accurate result when using TDDDM.

## VI. APPLICATIONS

### A. Size effect on frequency dependent thermal conductivity contribution

It is well known that the size effect cannot be avoided when NEMD method is used to calculate the thermal conductivity of materials. As analyzed above, size effect is inherent in our FDDDM simulations since it is based on the NEMD method. In our previous work [34], we have proved that the size effect for LJ Argon in the NEMD simulations can be ignored when the size of the system reaches about several hundreds of nanometers. Here the SW Si is chosen as the example to analyze the size effect in FDDDM. For SW Si, it has been demonstrated in literature that the thermal conductivity computed using NEMD can be strongly affected by the size of the model [18, 19, 47, 49]. The difference in thermal conductivity between NEMD and EMD method, which usually has small size effect, can be as large as a few times [18, 47, 52, 53]. The results of FDDDM for different total length of system are plotted in Fig. 3. From the left panel of Fig. 3, we can find that in low frequency region ($\omega <$ 6 THz), the frequency dependent thermal conductivity is found to be proportional to the excited transverse and longitude acoustic phonon modes, i.e. increasing as $\omega^2$. This phenomenon is also observed in Ref. [37]. In the frequency region from 6 to 12 THz, the contribution to the overall thermal conductivity decreases because the number of phonon modes is much less than that in the low frequency region (only longitude acoustic phonon mode in this range) [31]. In the high frequency region ($\omega >$ 12 THz), all phonons are optic. It is usually believed that the contribution of optic phonons can be ignored for SW Si. Here, in our simulations, the optic phonons contribute as little as 10% of the total thermal conductivity (right panel of Fig. 3), which is accordance with the common sense. It is also worth noting that more phonons in the low frequency region ($\omega <$ 6 THz), or equivalently phonons with longer MFP, appear in the FDDDM simulations with system size increasing (left panel of Fig. 3). Therefore, the contribution of the low frequency phonon modes rises with the increase of system size (right panel of Fig. 3). Furthermore, as discussed in the previous paper [34], the phonons in the low frequency region or with long MFP are truncated and then contribute little to the thermal conductivity. The similar phenomenon can be found here as well. However, it should be noted that the truncation of long MFP phonons does not mean that these phonons do not exist. Here, we recommend using the TDDDM to investigate the size effect in NEMD simulations, since we can observe the truncation of long MFP phonons clearly when using TDDDM. We also calculate the actual thermal conductivity using FDDDM (inset in Fig. 3). The value of our largest system shown in Fig. 3 is 275 W/mK. However, Goicochea *et. al.* [31] show the thermal conductivity of bulk Si is 380 W/mK using EMD with the same original SW potential. The large difference between our result and that of the reference should come from the size effect in NEMD simulations. To this end, we increase the length along the concerned direction to about 870 nm and the thermal conductivity increases to 300 W/mK. Using the linear extrapolation between $1/\kappa$ and $1/L$, we obtain the thermal conductivity of bulk Si to be 385 W/mK, which is in very good agreement with the result from Ref. [31]. One should also keep in mind that the length along the concerned direction used to do the linear extrapolation should be comparable to or longer than the dominant MFP in the system. One thing we have to mention here is that using the large samples (the four samples used to extrapolate the thermal conductivity of SW Silicon at 300 K is from ~115 nm to ~700 nm in our simulations) is necessary to obtain the accurate result.

## B. Temperature effect on spectral contribution to thermal conductivity

Another important issue is the temperature effect on the thermal conductivity. Four different system temperatures are used to investigate the temperature effects. The results of both LJ Argon (Fig. 4) and SW Si (Fig. 5) show that the anharmonic energy transfer, which is caused by the anharmonic phonon scattering, is enhanced with the increase of the temperature. For LJ Argon at 10K the phonon modes around 1.2 THz (the peak region in left panel of Fig. 4) can carry energy farther before scattering, as compared with the system at relatively higher temperatures such as 20 K, 30 K and 40K, because the anharmonic effect is weak at such low temperature. For the systems at temperature of 20 K, 30 K and 40 K, the above phonons still carry energy farther before scattering than other phonons. However, their contribution is decreased considerably with respect to that for 10 K. With the increase of temperature, the difference of thermal conductivity contribution between each phonon modes decreases. At relatively high temperature (40 K in our cases), the contribution of phonon modes is even almost equal to each other, i.e. the frequency dependent thermal conductivity contribution is very broad throughout the entire frequency range. It is also easy to find that, in high frequency region ($\omega > 1.7$ THz) the contribution of phonon modes almost remains unchanged at different temperatures. Phonons in this region can appear and annihilate easily, since these phonons usually have an extremely short lifetime. The reduction or improvement of relaxation time caused by the system temperature in this region is quite small, and therefore, leads to the constant contribution of phonons in this region. In addition, it is interesting to find that the thermal conductivity decreasing with temperature is nonlinear (right panel of Fig. 4), which is also observed in Ref. [37]. What is more, the actual thermal conductivity calculated using FDDDM for LJ Argon is 3.2 W/mK (10 K), 1.88 W/mK (20 K), 1.17 W/mK (30 K) and 0.88 W/mK (40 K). The four corresponding values are 4.0, 1.57, 0.903 and 0.574 W/mK computed using GK-MD [29]. The difference here should come from three aspects. Firstly, our FDDDM result is obtained by one simulation run. Using more simulation runs, we can obtain a more accurate result with an error bar. Secondly, there is still a weak size effect in our systems. Third, the parameters of the potential used in work of Ref. [29] have small difference with ours. In the work of Ref. [29], the authors use $\sigma = 0.34$ nm and $\varepsilon = 0.0104375$ eV, while in our simulations we choose $\sigma = 0.342$ nm and $\varepsilon = 0.0104233$ eV.

For SW Si, the profile of the thermal conductivity spectrum is scaled down for the entire frequency range with temperature increasing (left panel of Fig. 5). We can also see that the range of frequency decreases from 17.8 to 17.0 THz with temperature increasing, due to the well-known phonon softening effect at high temperatures. Since temperature has pronounced effect on both acoustic and optic phonons [31], it can be observed that the slope of the accumulative thermal conductivity with respect to frequency has significant change in both acoustic and optic phonon region with temperature increasing. It is also worth noting that, the reduction of the actual accumulative thermal conductivity for optic phonons due to the increase of the temperature is larger than that for acoustic phonons. In addition, we should mention that the actual thermal conductivity may be quite different from the bulk (infinite length) thermal conductivity at the same temperature conditions, since the length used in our model is quite short.

### C. Frequency dependent thermal conductivity of amorphous Si

The unique advantage of FDDDM prior to previous methods is the ability to deal with disordered systems. Other methods, e.g. SED and TDNMA, can only deal with the amorphous materials with assumptions, since one of the foundations involved in all of them is lattice dynamics. To certify this, in this section we perform FDDDM analysis to amorphous Si (a-Si). The method in Ref. [54] is used to generate the amorphous Si samples. It should be noticed that the amorphous material may have many different atomic configurations with nearly equivalent potential energies, and therefore, leading to the potential meta-stability during the MD simulations[55]. To overcome the meta-stability, all samples are annealed at a temperature of 3000 K for 5 ns. Moreover, the result discussed below is averaged by four independent runs. In order to verify the result of our new method, the phonon properties of a-Si are also computed by the Allen-Feldman (AF) theory, where the isotropic modal thermal conductivity which is a scalar can be calculated as[56]

$$\mathrm{K}_{AF}(\omega) = k_b D(\omega) \ , \tag{15}$$

where $D(\omega)$ is the AF diffuson diffusivity and can be calculated from[56]

$$D(\omega) = \frac{\pi V^2}{\hbar^2 \omega^2} \sum_{\omega' \neq \omega} |S_{\omega\omega'}|^2 \delta(\omega' - \omega) , \tag{16}$$

where $\delta$ is the Dirac delta function. The heat current operator $S_{\omega\omega'}$ measures the thermal coupling between vibrational modes $\omega$ and $\omega'$. The computing details of $S_{\omega\omega'}$ can be found in Ref. [56]. Finally, using Eqs. (14) and (15), the AF thermal conductivity per frequency can be obtained.

The results calculated using FDDDM and AF theory are compared in Fig. 6. Here, we also give the result of crystalline Si (c-Si) for comparison. General consistency is found between the results calculating with FDDDM and that of AF theory. Comparing with the results of FDDDM, the AF-predicted phonons have relatively high thermal conductivity contribution in the low-frequency region ($\omega$ < 8 THz). The reason is the broadening of the Dirac delta function, which can lead to the decreasing of diffusivities at intermediate and high frequencies [55]. In disordered materials, Larkin *et al.* [55] find that the relaxation time of each phonon in the middle frequency region (4~12 THz) is nearly in the same magnitude (Fig. 4 in Ref. [54]), which means the scattering of phonons in such materials is quite strong. Furthermore, we also plot the vibrational density of states (VDOS) of a-Si and c-Si, which shows that the distribution of phonons in a-Si is more uniform with respect to that of c-Si (inset of Fig. 6). Therefore, it is not surprising to find that the contribution of middle frequency phonons in a-Si is more uniform than that in c-Si (left panel of Fig. 6). We also observe that the phonons of a-Si in high frequency region contribute less than that of c-Si in the same region, or equivalently, phonons of a-Si in this region is more likely to scatter. Generally speaking, in the disordered materials, diffusons are the main heat carriers [55, 57]. Therefore, it is usual to treat the disordered materials as homogenous materials when the thermal conductivity of such material is concerned. Once again, we plot the actual thermal conductivity of the a-Si computed using FDDDM and AF theory. The results are 1.45 W/mK for FDDDM and 1.21 W/mK for AF theory. We think the difference should come from the fact that the propagating phonon mode which is not considered by AF theory can contribute to the total thermal conductivity to some extent. At the same time, Larkin *et al.*[55] give the thermal conductivity of a-Si as 2.0 W/mK using GK-MD and 1.2 W/mK using AF theory. The result obtained using GK-MD is larger than that by FDDDM and AF theory. This may be related to the propagating phonon modes, which deserves further study in the future.

## VII. CONCLUSIONS

In this paper we propose a new approach named frequency domain direct decomposition method based on non-equilibrium molecular dynamics simulation and time Fourier transform of atomistic heat current to quantitatively characterize frequency dependent thermal conductivity in solid materials. The validity of our new FDDDM is assessed by comparing the results of bulk LJ Argon and SW Si to TDNMA, which is based on the equilibrium molecular dynamics simulation. Some important parameters in FDDDM implementation and calculation are discussed explicitly. Comparing with previous methods such as SED, TDNMA, and TDDDM, the new FDDDM approach has unique features and power in the following aspects: (1) Due to the inherent nature that FDDDM is based on NEMD, we can use FDDDM to explicitly investigate the size effect of frequency dependent thermal conductivity. Generally speaking, more phonons in the low frequency region, or equivalently phonons with long MFP, are present when the system size increases. (2) There is no assumption of the system temperature in our theoretic model, therefore, in principle the FDDDM can be used to observe the temperature effect of phonon mode thermal conductivity directly. Anharmonic energy transfer is found to increase with temperature increasing, which is well known but proves that our FDDDM reveals the correct physical picture of the phonon-phonon scattering. (3) The FDDDM can be directly used to study the phonon behavior of disordered systems. The amorphous Si is chosen as a benchmark system to prove the unique advantage of our FDDDM method. The contribution of phonons are found to be quite uniform, since the relaxation time of phonons in amorphous materials is substantially short and therefore the scattering is strong as compared with the crystalline materials. Moreover, the framework of our FDDDM approach is expected to be straightforwardly degenerated to some special situations such as interface problem, which is undergoing currently. The present method for determining the contributions of different vibrational frequencies to the overall thermal conductivity is expected to be quite useful in thermal engineering applications, such as realizing high efficiency thermoelectrics by manipulating phonons in nanostructures and rationally designing multi-layers for advanced thermal management.

## Acknowledgements

Y.Z. thanks Dr. Xiaoliang Zhang (RWTH Aachen University) and Mr. Kimmo Sääskilahti (Aalto University) for useful discussions. Y.Z. also thanks Dr. Tao Wang (Ruhr University Bochum) for his help in calculating vibration density of states of crystalline Si and Dr. Davide Donadio (Max Planck Institute for Polymer Research) for his code of AF Theory. Simulations were performed with computing resources granted by the Jülich Aachen Research Alliance-High Performance Computing (JARA-HPC) from RWTH Aachen University under Project No. jara0127.

## Appendix A

In order to prove the accuracy of Eq. (9), we plot the atomic displacement along three directions for LJ Argon at 40 K and SW Si at 600 K. From FIG. A1 we can find that the maximum displacement with respect to their equilibrium position for both LJ Argon and SW Si is about 0.012$a$, where $a$ is their respective lattice constant. Then, we know the maximum $(\mathbf{u}_{s_\alpha} - \mathbf{u}_{s_\beta})$ is about 0.024$a$, which is much smaller than $a$. Thus, the assumption in Eq. (9) is appropriate.

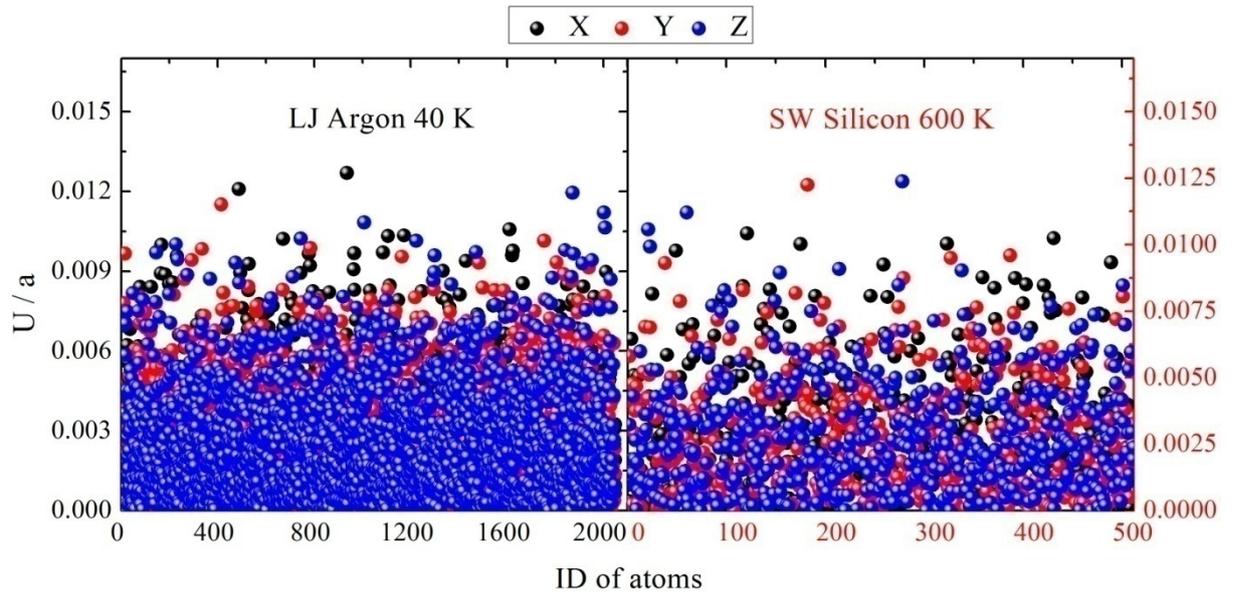

FIG. A1. The absolute value of atomic displacement along three directions for (left) LJ Argon at 40 K and (right) SW Si at 600K. The displacement ($U$) is normalized by the respective lattice constant ($a$).

# References

1 J. H. Lienhard and J. Lienhard, *A heat transfer textbook* (Phlogiston Press, Cambridge, 2000).

2 T. M. Tritt, *Theory, properties and applications, Physics of Solid and Liquids,* (Kluwer Academic/Plenum Publishers, New York, 2004).

3 D. G. Cahill, P. V. Braun, G. Chen, D. R. Clarke, S. Fan, K. E. Goodson, P. Keblinski, W. P. King, G. D. Mahan and A. Majumdar, Appl. Phys. Rev. **1**, 11305 (2014).

4 P. G. Klemens, Proc. R. Soc. London, Ser. A **68**, 1113 (1955).

5 A. A. Maradudin and A. E. Fein, Phys. Rev. **128**, 2589 (1962).

6 A. Debernardi, S. Baroni and E. Molinari, Phys. Rev. Lett. **75**, 1819 (1995).

7 G. Deinzer, G. Birner and D. Strauch, Phys. Rev. B **67**, 144304 (2003).

8 P. G. Klemens, Proc. R. Soc. London, Ser. A **208**, 108 (1951).

9 F. P. Incropera, *Fundamentals of heat and mass transfer* (John Wiley & Sons, 2011)

10 T. Feng and X. Ruan, J. Nanomater. **2014**, 206370 (2014).

11 A. McGaughey and J. M. Lakin, Annual. Rev. Heat Transfer **17**, 49 (2014).

12 Y. Chalopin, A. Rajabpour, H. Han, Y. Ni and S. Volz, Annual. Rev. Heat Transfer **17**, 147 (2014).

13 J. Wang, J. Wang and J. T. Lü, Eur. Phys. J. B **62**, 381 (2008).

14 D. A. Broido, M. Malorny, G. Birner, N. Mingo and D. A. Stewart, Appl. Phys. Lett. **91**, 231922 (2007).

15 W. Li, J. Carrete, N. A. Katcho and N. Mingo, Comp. Phys. Comm. **185**, 1747 (2014).

16 A. Chernatynskiy and S. R. Phillpot, Comp. Phys. Comm. **192**, 196 (2015).

17 R. Kubo, Prog. Phys. **29**, 255 (1966).

18 S. G. Volz and G. Chen, Phys. Rev. B **61**, 2651 (2000).

19 P. K. Schelling, S. R. Phillpot and P. Keblinski, Phys. Rev. B **65**, 144306 (2002).

20 C. Melis, R. Dettori, S. Vandermeulen and L. Colombo, Eur. Phys. J. B **87**, 1 (2014).

21 J. Shiomi, Annual. Rev. Heat Transfer **17**, 177 (2014).

22 J. A. Thomas, J. E. Turney, R. M. Iutzi, C. H. Amon and A. J. McGaughey, Phys. Rev. B **81**, 81411 (2010).

23 J. M. Larkin, J. E. Turney, A. D. Massicotte, C. H. Amon and A. McGaughey, J. Comp. Theor. Nanos. **11**, 249 (2014).

24 B. Qiu and X. Ruan, Appl. Phys. Lett. **19**, 100, (2012).

25 B. Qiu and X. Ruan, Appl. Phys. Lett. **100**, 193101 (2012).

26 A. S. Henry and G. Chen, J. Comput. Theor. Nanosci. **5**, 141 (2008).

27 B. Qiu, H. Bao, G. Zhang, Y. Wu and X. Ruan, Comp. Mater. Sci. **53**, 278 (2012).

28 A. J. Ladd, B. Moran and W. G. Hoover, Phys. Rev. B **34**, 5058 (1986).

29 A. McGaughey and M. Kaviany, Int. J. Heat Mass Transf. **47**, 1783 (2004).

30 A. J. McGaughey and M. Kaviany, Phys. Rev. B **69**, 94303 (2004).

31 J. V. Goicochea, M. Madrid and C. Amon, J. Heat Transf. **132**, 012401 (2010).

32 T. Hori, T. Shiga and J. Shiomi, J. Appl. Phys. **113**, 203514 (2013).

33 J. M. Larkin and A. J. McGaughey, J. Appl. Phys. **114**, 23507 (2013).

34 Y. Zhou, X. Zhang and M. Hu, arXiv:1502.06776.

35 K. Gordiz and A. Henry, arXiv:1407.6410.

36 Y. He, D. Donadio and G. Galli, Appl. Phys. Lett. **98**, 144101 (2011).

37 K. Sääskilahti, J. Oksanen, J. Tulkki and S. Volz, Phys. Rev. B **90**, 134312 (2014).

38 Y. Chalopin and S. Volz, Appl. Phys. Lett. **103**, 51602 (2013).

39 Y. Chalopin, K. Esfarjani, A. Henry, S. Volz and G. Chen, Phys. Rev. B **85**, 195302 (2012).

40 J. H. Irving and J. G. Kirkwood, J. Chem. Phys. **18**, 817 (1950).

41 D. Torii, T. Nakano and T. Ohara, J. Chem. Phys. **128**, 44504 (2008)

42 T. Ohara, J. Chem. Phys. **111**, 9667 (1999).

43 It is true that Eq. (7) in this paper is another form of Eq. (1) in our first paper in the series. One can write the detailed expression of virial stress in Eq. (1), then Eq. (1) should be consistent with Eq. (7). However, the authors cannot give rigorous proof for this currently.

44 J. Li, Ph.D. thesis, *http://li.mit.edu/Archive/CourseWork/Ju_Li/Thesis/*.

45 G. P. Srivastava, *The physics of phonons* (CRC Press, 1990).

46 J. E. Turney, E. S. Landry, A. McGaughey and C. H. Amon, Phys. Rev. B **79**, 64301 (2009).

47 D. P. Sellan, E. S. Landry, J. E. Turney, A. McGaughey and C. H. Amon, Phys. Rev. B **81**, 214305 (2010).

48 J. E. Turney, A. McGaughey and C. H. Amon, J. Appl. Phys. **107**, 24317 (2010).

49 D. P. Sellan, J. E. Turney, A. McGaughey and C. H. Amon, J. Appl. Phys. **108**, 113524 (2010).

50 M. S. Bartlett, Nature **161**, 686 (1948).

51 S. Plimpton, J. Comp. Phys. **117**, 1 (1995).

52 X. Zhang, H. Xie, M. Hu, H. Bao, S. Yue, G. Qin and G. Su, Phys. Rev. B **89**, 54310 (2014).

53 M. H. Khadem and A. P. Wemhoff, Comp. Mater. Sci. **69**, 428 (2013).

54 A. France-Lanord, E. Blandre, T. Albaret, S. Merabia, D. Lacroix and K. Termentzidis, J. Phys. **26**, 055011 (2014).

55 J. M. Larkin and A. J. McGaughey, Phys. Rev. B **89**, 144303 (2014).

56 P. B. Allen and J. L. Feldman, Phys. Rev. B **48**, 12581 (1993).

57 G. C. Sosso, D. Donadio, S. Caravati, J. Behler and M. Bernasconi, Phys. Rev. B **86**, 104301 (2012).

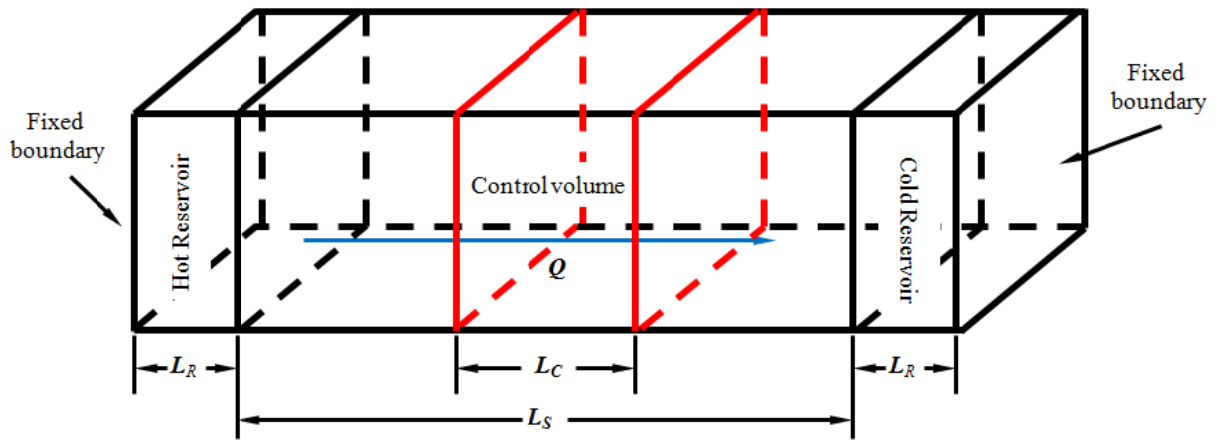

FIG. 1. (Color online) Schematic of FDDDM (NEMD) simulation. $L_S$, $L_C$, and $L_R$ is the total length of the system, the length of the control volume, and the length of thermostats, respectively. $Q$ stands for the heat current in the system.

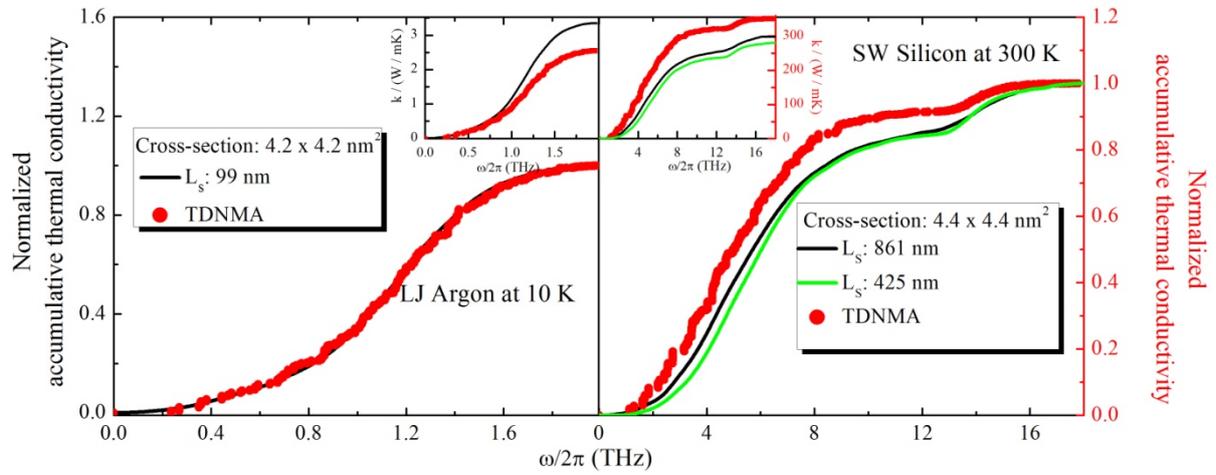

FIG. 2. (Color online) Accumulative thermal conductivity as a function of mode frequency for the case of (left) LJ Argon and (right) SW Si calculated using TDNMA and FDDDM methods. The accumulative thermal conductivity is normalized by the overall thermal conductivity predicted by the respective method. (Inset) The actual accumulative thermal conductivity vs. frequency.

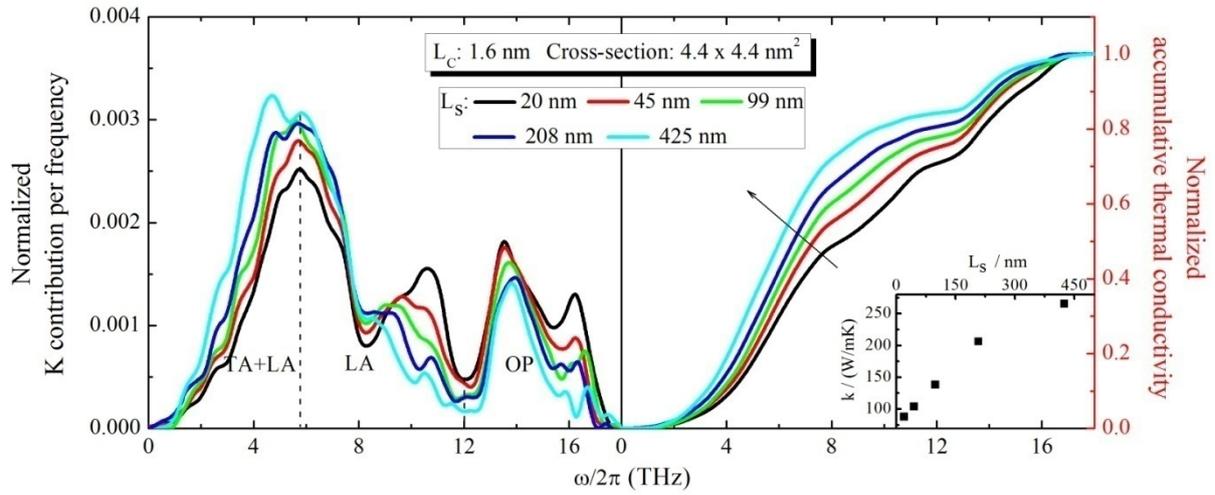

FIG. 3. (Color online) (Left) Normalized spectral contribution to thermal conductivity and (right) corresponding normalized accumulative thermal conductivity as a function of mode frequency for the case of SW Si with different system size computed using FDDDM. The dashed lines and the arrow are guide for the eyes. (Inset) Actual thermal conductivity of SW Si vs. system length.

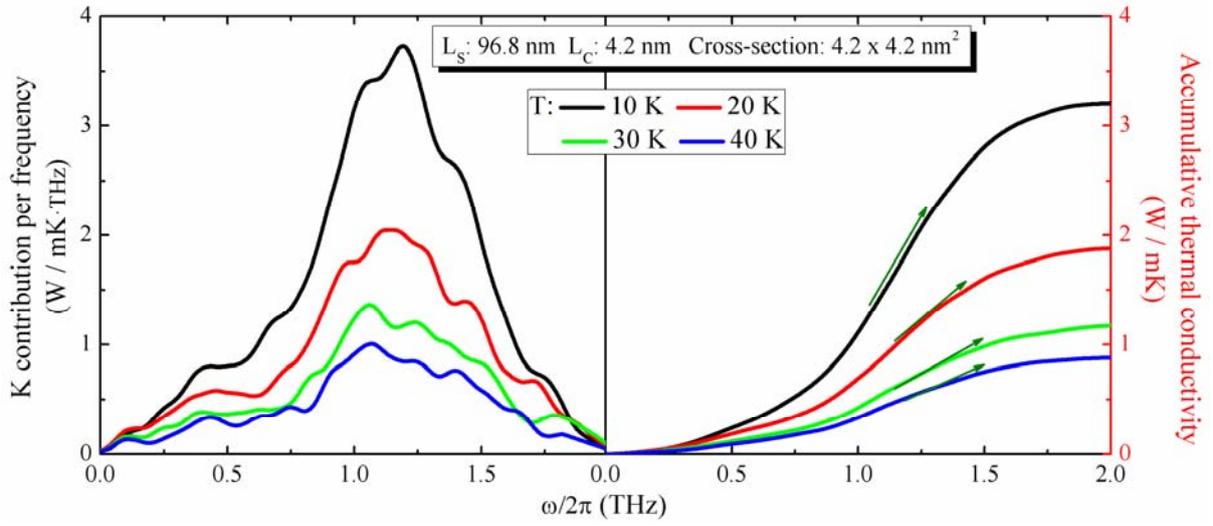

FIG. 4. (Color online) (Left) Spectral contribution to thermal conductivity and (right) corresponding accumulative thermal conductivity as a function of mode frequency for the cases of LJ Argon with different system temperature. The arrows are guide for the eyes.

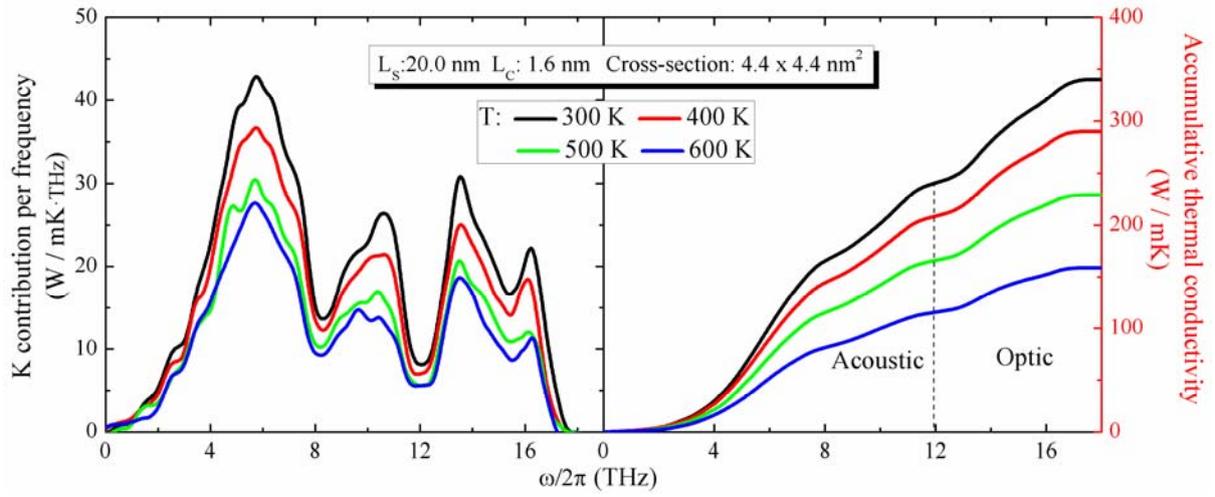

FIG. 5. (Color online) Same figure as Fig. 4 for the case of SW Si. The dashed line is guide for the eyes.

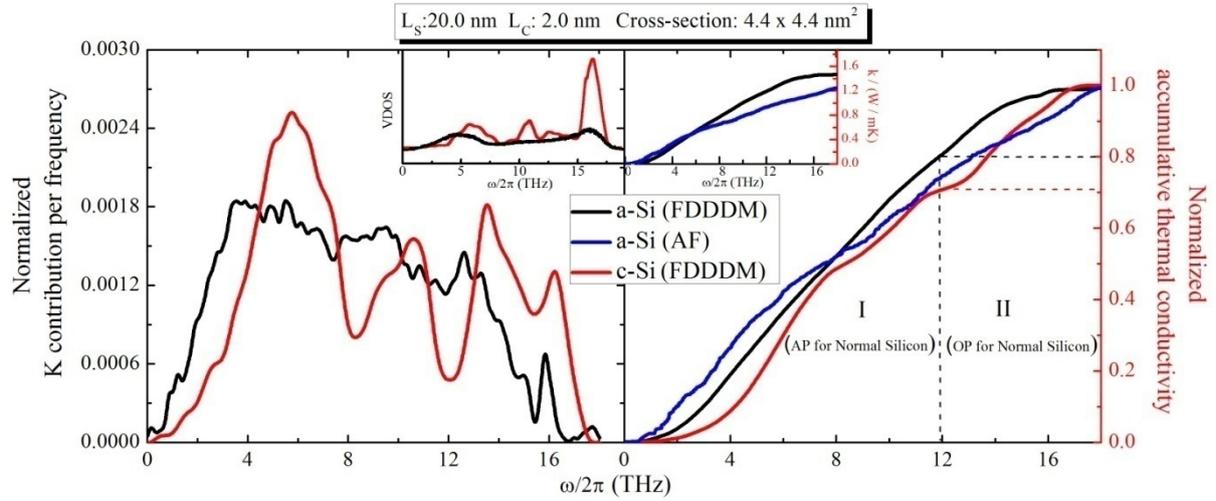

FIG. 6. (Color online) (Left) Normalized spectral contribution to thermal conductivity and (right) corresponding normalized accumulative thermal conductivity as a function of mode frequency for amorphous Si computed using FDDDM and AF theory. The dashed lines are guide for the eyes. (Left inset) Comparison of VDOS between amorphous Si and crystalline Si. (Right inset) The actual accumulative thermal conductivity of c-Si and a-Si vs. frequency.